\documentclass[aip,reprint]{revtex4-1}
\draft 
\usepackage{algpseudocode, algorithm, graphics, amsmath, physics}

\begin{document}
\title[]{Detection of temporal fluctuation in superconducting qubits for quantum error mitigation}
\author{Yuta Hirasaki}
 \affiliation{ 
Department of Applied Physics, The University of Tokyo, Tokyo 113-8656, Japan.
}%
\author{Shunsuke Daimon}%
 \email[Author to whom correspondence should be addressed: ]{daimon.shunsuke@qst.go.jp}
 \affiliation{ 
Department of Applied Physics, The University of Tokyo, Tokyo 113-8656, Japan.
}%
 \affiliation{ 
Quantum Materials and Applications Research Center, National Institutes for Quantum Science and Technology (QST), Tokyo 152-8550, Japan.
}%
\author{Toshinari Itoko}
\author{Naoki Kanazawa}
\affiliation{%
IBM Quantum, IBM Research-Tokyo, 19-21 Nihonbashi Hakozaki-cho, Chuo-ku, Tokyo, 103-8510, Japan.
}%
\author{Eiji Saitoh}
 \affiliation{ 
Department of Applied Physics, The University of Tokyo, Tokyo 113-8656, Japan.
}%
\affiliation{ Institute for AI and Beyond, The University of Tokyo, Tokyo 113-8656, Japan.}
\affiliation{WPI Advanced Institute for Materials Research, Tohoku University, Sendai 980-8577, Japan.}
\affiliation{Institute for Materials Research, Tohoku University, Sendai 980-8577, Japan.}

\date{\today}

\begin{abstract}
We have investigated instability of a superconducting quantum computer by continuously monitoring the qubit output. We found that qubits exhibit a step-like change in the error rates. This change is repeatedly observed, and each step persists for several minutes. By analyzing the correlation between the increased errors and anomalous variance of the output, we demonstrate quantum error mitigation based on post-selection. Numerical analysis on the proposed method was also conducted.
\end{abstract}


\maketitle 
Over the last few decades, there has been a growing trend towards developing quantum computers and advances in quantum engineering technologies are overwhelming \cite{ladd2010quantum, neill2018blueprint}. Among diverse materials or artificial atoms proposed to serve as quantum bits (qubits), superconducting qubits\cite{nakamura1999coherent, krantz2019quantum} are one of the most promising candidates. A number of studies have been conducted to improve the performance of superconducting qubits and several breakthroughs have been achieved\cite{kjaergaard2020superconducting, schreier2008suppressing, koch2007charge}. Nevertheless, even the state-of-the-art qubits unpredictably interact with the surrounding environments and suffer from noise during computation, which places a critical limit on their computational abilities\cite{unruh1995maintaining, preskill2018quantum, bharti2022noisy}.

Several attempts have been made to identify microscopic pictures of unexpected interactions and improve the device's performance \cite{muller2019towards, martinis2021saving, constantin2007microscopic, martinis2005decoherence}. Recent evidence suggests that superconducting qubits exhibit a temporal change in their coherence times under a continuous measurement \cite{klimov2018fluctuations, muller2015interacting, carroll2022dynamics, bylander2011noise, thorbeck2022tls, vepsalainen2020impact, de2020two, shnirman2005low}.
Qubit instability poses a serious threat to quantum computers. A sudden decrease in the qubit lifetime can temporarily degrade the device's performance.
In addition, most of the current quantum error mitigation (QEM) techniques \cite{van2023probabilistic, kim2023scalable, temme2017error, kandala2019error} are unable to mitigate time dependent noise\cite{schultz2022impact}, and a temporal change in decoherence calls for re-learning of a noise model or developing more sophisticated QEM techniques. Therefore, it is imperative to investigate the dynamics of a superconducting qubit system and assess its stability.

In this paper, we report a temporal change in the qubit errors in a superconducting quantum computer. We also developed an anomaly detection method for a temporal change in errors.

All the experiments were performed on \texttt{ibm\_kawasaki}, which is one of the IBM Quantum systems. This quantum computer has 27 transmon qubits and the readout assignment errors are around $1\%$ on average. The energy relaxation times of the qubits are approximately $1.2\times 10^2\;\mathrm{\mu s}$ on average, with the phase damping times around $1.2\times 10^2\;\mathrm{\mu s}$. 

We iterate a same quantum circuit and a subsequent measurement for $L$ times at a sampling rate of several hundred microseconds. As a result, we obtain a binary sequence $\mathbf{X} \in \{0, 1\}^L$. To estimate the qubit output fluctuations, we transform a subsequence of $\mathbf{X}$ with size $N$ into a fluctuation indicator $S$, which is defined by
\begin{align}\label{eq1}
  S = {\frac{1}{m-1}\sum_{j = 1}^{m}(Y_j - \overline{Y})^2}\left/{\frac{\overline{Y} ( 1 - \overline{Y})}{n}}\right.,
\end{align}
where $Y_j = \frac{1}{n}\sum_{i = (j-1)n + 1}^{jn}X_i$, and $\overline{Y} = \frac{1}{m}\sum_{j = 1}^mY_j$ with some integers $n$ and $m$ that satisfy the condition $N = nm\ll L$. In the experiments below, we obtain a time series of $S$ from the entire sequence $\mathbf{X}\in\{0, 1\}^L$ using the following procedure. We first take the average of every $n$ data to obtain a time series $\mathbf{Y}$ with the length $M = \left\lfloor\frac{L}{n}\right\rfloor$. We then calculate the time series $\mathbf{S}$ from $\mathbf{Y}$ by applying a sliding window of size $m$, and thus the length of $\mathbf{S}$ is given by $l = M- m + 1$.

The indicator $S$ is introduced based on the following background. From the Born's rule, the measurement outcome  $X_i$ in the $i$-th measurement is a random variable whose distribution is given by the binomial distribution $B(1, P_1)$, where $P_1$ denotes the probability of measuring the excited state. The average $Y_j$ is also a random variable whose probability distribution is determined by the binomial distribution $B(n, P_1)$. Thus, the expectation value of the sample mean $\overline{Y} = \frac{1}{m}\sum_{j = 1}^mY_j$ is equal to $P_1$, and that of the unbiased sample variance $V_{\mathrm{samp}} = \frac{1}{m-1}\sum_{j = 1}^m(Y_j-\overline{Y})^2$ is equal to $\frac{P_1(1 - P_1)}{n}$. Since $P_1$ is unknown, we estimate the expected variance with $V_\mathrm{bi} = \frac{\overline{Y}(1-\overline{Y})}{n}$, and $S$ is given by the ratio of $V_\mathrm{samp}$ and $V_\mathrm{bi}$ in Eq. (\ref{eq1}). Intuitively, $S$ quantifies the extent to which the sample variance deviates from what is expected under the assumption that $\{X_i\}_i$ are generated from an identical binomial distribution. $S$ can be used to detect a temporal change in qubit errors and exclude abnormal outcomes in quantum computing as discussed later.

Note that $S$ is a random variable obtained from the random variables $X_1, X_2, \dots, X_N$ and $S$ takes several values with different probabilities. The probability distribution of $S$ is well described by the chi-squared distribution with $(m-1)$ degrees of freedom and the mean of $S$ is given by $1$ with the variance $\sigma^2 = \frac{2}{m-1}$, whose rigorous derivation is provided in the latter part of this letter. Thus, when we calculate $S$ from an experimental result (for clarity we represent the experimental value as $S_\mathrm{exp}$ and use $S_\mathrm{theo}$ when we describe a stochastic characteristic of $S$), $S_\mathrm{exp}$ should spread randomly around $1$ with the statistical fluctuation $\sigma = \sqrt\frac{2}{m-1}$. If $S_\mathrm{exp}$ significantly deviates from the probabilistic behavior of $S_\mathrm{theo}$, we reject the hypothesis that the  binary data $X_1, X_2,\dots ,X_N$ are generated from an identical binomial distribution $B(1, P_1)$ and the data are classified as anomalous in our QEM method.

\begin{figure}
\includegraphics{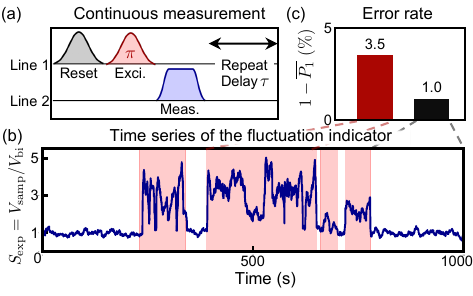}
\caption{
(a) The pulse sequence of the experiment. The repeat delay $\tau$ is approximately $6\times10^2 \mathrm{\mu s}$. (b) The time series of our indicator $S_\mathrm{exp}$ with $n = m = 128$. (c) The error rates in two time periods. The red bar shows the value of $1 - \overline{P_1}$ in the red-shaded region as indicated by the red dashed line. The black bar shows the error rate in the normal (not-shaded) time period, as indicated by the gray dashed line.
}
\label{f1}
\end{figure}

First, we performed a one-qubit continuous measurement on the IBM quantum processor. The pulse sequence is depicted in Fig. \ref{f1}(a). The qubit is initialized to the ground state with the reset pulse, excited with the $\pi$ pulse, and then measured. We repeated this pulse sequence for 1000 seconds with the repeat delay time $\tau\approx 6\times 10^2 \;\mathrm{\mu s}$ to record normal and abnormal behavior in a single set of experimental data. The time series of $S_\mathrm{exp}$ defined by Eq. (\ref{eq1}) was calculated from the obtained outcomes with the parameters $n = m = 128$ and $L = 1787904$.

Figure \ref{f1}(b) illustrates the time series of $S_\mathrm{exp}$. The value of $S_\mathrm{exp}$ remains almost constant for the first 230 s. This behavior is consistent with the fact that the expectation value of $S_\mathrm{theo}$ is equal to $1$ with the standard deviation $\sigma \approx 0.125$. In the next moment, however, $S_\mathrm{exp}$ abruptly increases to approximately 4 [see the red band in Fig. \ref{f1}(b)], which is 24 standard deviations above the mean, and this cannot be explained in terms of the statistical error. This increase persists for 110 seconds, and sharp switching behavior is repeatedly observed in the rest of the record as visualized by the four red bands in Fig. \ref{f1}(b). This phenomenon is observed repeatedly in other experiments on \texttt{ibm\_kawasaki}.

Figure \ref{f1}(c) compares the error rates in two time periods. The red bar represents $1 - \overline{P_1}$ in the time period from 430 s to 720 s, while the black bar shows that from 870 s to 1000 s, where $\overline{P_1}$ denotes the average of the binary outcomes and should be $1$ in the absence of errors. The temporal increase in $S_\mathrm{exp}$ appears to be closely related to a temporal increase in errors. This correlation between $S_\mathrm{exp}$ and errors suggests that we can reduce errors by classifying obtained outcomes based on the values of $S_\mathrm{exp}$ and eliminating the anomalous outcomes.

Based on this, we propose a QEM technique based on post-selection (or we also call it an anomaly detection). We first compute the time series $\mathbf{S}_\mathrm{exp}$ from an obtained binary sequence $\mathbf{X}$. Then, we compare each element of $\mathbf{S}_\mathrm{exp}$ against a threshold value $S_\mathrm{thre}$. If an element exceeds the threshold, we label the corresponding subsequence of $\mathbf{X}$ as anomalous and segregate it from the remaining sequence. The critical value is determined based on the $p$-value in the detection and here we employ $S_\mathrm{thre} = 1.5$, which corresponds to the $p$-value of $0.006334\%$. This method can be easily extended to multi-qubit computations by computing the time series of $S_\mathrm{exp}$ for each qubit individually.

\begin{figure}
  \includegraphics{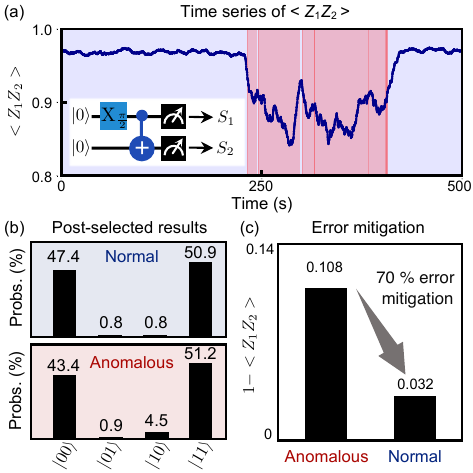}
  \caption{(a) The time series of $\ev*{Z_1Z_2}$, with the sampling interval approximately $6\times 10^2\;\mathrm{\mu s}$. We prepared the Bell state $\frac{\ket{00} + \ket{11}}{\sqrt{2}}$ with the $\pi/2$ pulse and the C-NOT gate. The anomaly detection was performed on the measured two qubits. The data (or the time periods) that are classified as the anomalous (normal) state are colored with red (blue). (b) The probability distributions obtained from the classified data. The top (bottom) panel presents the probability distribution computed from the normal (anomalous) state. (c) The values of $1-\ev*{Z_1Z_2}$ calculated from the post-selected distributions. The column on the left (right) represents the value of the anomalous (normal) state.
  }
  \label{f2}
\end{figure}

We performed a Bell state measurement to demonstrate the proposed QEM as illustrated in Fig. \ref{f2}. We obtain two binary sequences from two qubits and calculated the time series of $S_\mathrm{exp}$ from the two sequences individually. For each time window with size $N$, we calculate $S_1$ and $S_2$ from the two binary subsequences by Eq. (\ref{eq1}). If either $S_1$ or $S_2$ exceeds the threshold value $S_\mathrm{thre} = 1.5$, the corresponding two binary subsequences are labeled as anomalous and labeled as normal otherwise. The time series of $\ev*{Z_1Z_2}$ is depicted in Fig. \ref{f2}(a), where $\ev*{Z_1Z_2}$ denotes the expectation value of the observable $Z_1Z_2$, and it is calculated from the two binary sequences with the same window. $\ev*{Z_1Z_2}$ should be $1$ in the absence of errors. The red colored region represents the time periods labeled as anomalous based on $S_\mathrm{exp}$ and the blue represents the normal state. $\ev*{Z_1Z_2}$ exhibits a great decrease to around $0.85$ in the anomalous time period [the red band in Fig. \ref{f2}(a)], while it shows little fluctuation around $0.97$ in the normal time periods. 

We obtain two histograms from the normal and anomalous outcomes as depicted in Fig. \ref{f2}(b). The probabilities of measuring the four states, $\ket{00},\ket{01}, \ket{10}$, and $\ket{11}$, are visualized by the black bars in Fig. \ref{f2}(b). The top panel shows the probability distribution calculated from the data classified as the normal state [colored blue in Fig. \ref{f2}(a)], while the one at the bottom depicts that from the anomalous state (colored red). The probability distribution of the anomalous state exhibits a prominent peak in the $\ket{10}$ state. We compare the values of $1 - \ev*{Z_1Z_2}$ obtained from the two categorized data as shown in Fig. \ref{f2}(c). This means that our method successfully removes the abnormal data and improves the fidelity in estimating the expectation value of a physical observable.

\begin{figure}
  \includegraphics{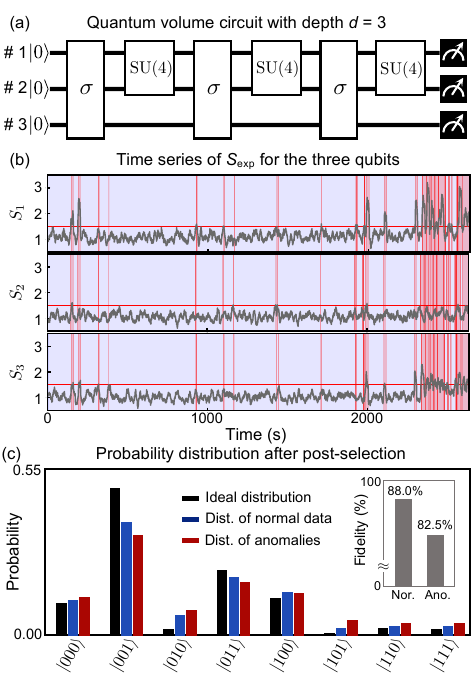}
  \caption{ 
  (a) The circuit consists of three qubits and three random unitary transformations, each of which is composed of a random permutation $\sigma$ and random unitary transformation on the first two qubits.
  (b) The time series of the fluctuation indicators for the three qubits. The red-colored regions represent the time periods when either $S_1, S_2$, or $S_3$ exceeds the threshold value, while the blue-shaded regions denote the normal time periods. Here we employ $n = m = 128$, and the sampling interval is approximately $3\times 10^2\;\mathrm{\mu s}$. We set the critical value $S_\mathrm{thre} = 1.5$.
  (c) The black bars represent the ideal probability distribution of the executed quantum circuit. The blue (red) column reflects the probability distribution of the normal (anomalous). The fidelities calculated based on the Hellinger distance are illustrated in the inset.  
  }
  \label{f3}
\end{figure}

We then benchmarked the proposed protocol in a quantum volume circuit \cite{cross2019validating} as an example of sampler tasks, in which we measure the probability distributions of the final quantum states. The result is shown in Fig. \ref{f3}. The circuit comprises three qubits and the qubits are measured after three layers of operation as shown in Fig. \ref{f3}(a). Each layer is characterized by sampling a random permutation and then applying a random unitary transformation to the first two qubits. 

We compute the time series of $S_\mathrm{exp}$ for the three qubits and classify the outcomes into the anomalous and normal state data as illustrated in Fig. \ref{f3}(b). The blue regions represent the outcomes classified as normal, while the red corresponds to the anomalous. We obtain two probability distributions from the two categorized experimental data and compare them with the ideal distribution (the black bars) as depicted in Fig. \ref{f3}(c). The distribution derived from the normal data is overall closer to the ideal distribution, demonstrating a $5.5\%$ improvement in the Hellinger fidelity\cite{le2000asymptotics}.

We note that in our setup the circuit outcomes have been recorded for a sufficiently long time to investigate the time variation of $S_\mathrm{exp}$. However, our mitigation technique can be applied at a moderate sampling overhead of tens of thousands shots, which is readily available with IBM Quantum processors.

Finally, we perform a theoretical analysis on the probability distribution of $S_\mathrm{theo}$ introduced in Eq. (\ref{eq1}). Note that the $i$-th measurement outcome $X_i$ is given by a random variable following the Bernoulli distribution $B(1, p_i)$, where $p_i$ is the probability of measuring the excited state in the $i$-th measurement. Here we make two fundamental assumptions, namely, $p_i$ is a constant $P_1$, and $\{X_i\}_i$ independently obey the identical Bernoulli distribution. Under these assumptions, it analytically follows that the random variables $nY_j = \sum_{i = nj + 1}^{(n + 1)j}X_i$ independently obey the binomial distribution $B(n, P_1)$ and the variance of $\{Y_j\}_j$ is given by $\frac{P_1(1 - P_1)}{n}$. Since $n$ is sufficiently large (in the experiments $n = 128$), we can apply the central limit theorem and approximate the probability distribution of $\{Y_j\}_j$ with a Gaussian distribution. Then we express $S_\mathrm{theo}$ in Eq. (\ref{eq1}) in terms of new random variables $\{Z_j\}_j$ defined by $Z_j = \frac{Y_j - P_1}{\sqrt{\frac{P_1(1 - P_1)}{n}}}$, which independently obey the standard normal distribution $\mathcal{N}(0, 1)$, where $\mathcal{N}(\mu, \sigma^2)$ denotes a Gaussian distribution with the mean $\mu$ and the variance $\sigma^2$. The expression of $S_\mathrm{theo}$ is given by
\begin{align}
\label{eq3}
  S_\mathrm{theo} = \frac{\frac{1}{m-1}\sum_{j = 1}^m {(Z_j-\overline{Z})^2}}{\left( \frac{\overline{Z}}{\sqrt{n}} + \sqrt{\frac{P_1}{1 - P_1}} \right)\left( -\frac{\overline{Z}}{\sqrt{n}} + \sqrt{\frac{1 - P_1}{P_1}} \right)},
\end{align}
where $\overline{Z} = \frac{1}{m}\sum_{j = 1}^m Z_j\sim \mathcal{N}\left( 0, \frac{1}{m} \right)$. $\frac{\overline{Z}}{\sqrt{n}}$ takes values of order $\frac{1}{\sqrt{nm}}$ with high probability, and thus, when $\frac{1}{\sqrt{nm}}$ is much smaller than $\sqrt{\frac{P_1}{1 - P_1}}$ and $\sqrt{\frac{1 - P_1}{P_1}}$, $\frac{\overline{Z}}{\sqrt{n}}$ is negligible compared to $\sqrt{\frac{1 -P_1}{P_1}}$ and $\sqrt{\frac{P_1}{1 - P_1}}$ with a high likelihood. As a result, Eq. (\ref{eq3}) reduces to 
\begin{align}\label{eq4}
  S_\mathrm{theo}\approx \tilde{S} \equiv \frac{1}{m-1}\sum_{j = 1}^m(Z_j-\overline{Z})^2
\end{align}
$\sum_{j = 1}^m(Z_j - \overline{Z})^2$ obeys the chi-squared distribution with $(m - 1)$ degrees of freedom\cite{cochran1934distribution} and therefore the statistical characteristic of $\tilde{S}$ is analytically derived. In particular, the mean of $\tilde{S}$ is $\mu = 1$ and the variance is $\sigma^2 = \frac{2}{m - 1}$, which is independent of $P_1$. This fact suggests that we can use the same threshold for anomaly detection in practical quantum computation where $P_1$ (or the measured quantum state) is unknown. The condition $\sqrt\frac{1 - P_1}{P_1}, \sqrt\frac{P_1}{1-P_1}\gg \frac{1}{\sqrt{nm}}$ is satisfied in most of our experiments since we use $n = m = 128$, and the inequality $0.01\leq P_1\leq 0.99$ holds due to the $1\%$ readout assignment errors.

\begin{figure}
  \includegraphics{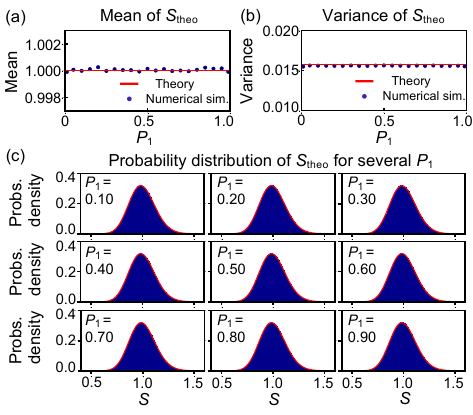}
  \caption{
  (a) The mean values of $S_\mathrm{theo}$ calculated from a Monte-Carlo simulation for several $P_1$ values (the blue dots) and the mean of $\tilde{S}$ (which is equal to $1$) derived from the theoretical result (the red line). (b) The variances of $S_\mathrm{theo}$ computed from the numerical simulation (the blue dots) and that of $\tilde{S}$ from the theoretical examination $\sigma^2 = \frac{2}{m - 1}$ (the red line). Here, we set $n = m = 128$. (c) Comparison of the probability density functions obtained from the numerical simulation (the blue histograms) and the chi-squared distribution (the red lines).
  }
  \label{f4}
\end{figure}

We then performed a Monte-Carlo simulation to support the validity of the discussions above, and the result is illustrated in Fig. \ref{f4}. We numerically prepared $100,000$ samples of $S_\mathrm{theo}$ for each of $P_1$ values we chose and compared the distributions of $S_\mathrm{theo}$ with those of $\tilde{S}$. The sample means of $S_\mathrm{theo}$ for several $P_1$ values (the blue dots) and the expectation value of $\tilde{S}$ ($\ev*{\tilde{S}} = 1$) (the red line) are depicted in Fig. \ref{f4}(a), while Fig. \ref{f4}(b) compares the variance of $S_\mathrm{theo}$ and $\tilde{S}$. The result provides a close similarity between the numerical and theoretical analysis for all the $P_1$ values. The probability density functions generated from the Monte-Carlo simulation are presented with the blue histograms in Fig. \ref{f4}(c) for several $P_1$ values. The red lines show the functions calculated theoretically, showing a good agreement with the numerical histograms. 

In conclusion, we have investigated a temporal change in fluctuations in superconducting qubits by developing a statistic that quantifies the qubit stability. The measured temporal change is closely related to a temporal increase in errors, and we have demonstrated QEM by analyzing the correlation of the fluctuation. Furthermore, we have conducted an analytical study on the QEM method, and performed a numerical simulation to verify the result.

\begin{acknowledgments}
This work was supported by CREST (Nos. JPMJCR20C1, JPMJCR20T2) from JST, Japan; Grant-in-Aid for Scientific Research (S) (No. JP19H05600), Grant-in-Aid for Transformative Research Areas (No. JP22H05114) from JSPS KAKENHI, Japan. 
This work is partly supported by IBM-Utokyo lab.
\end{acknowledgments}

\section*{Data Availability Statement}
The data that support the findings of this study are available from the corresponding author upon reasonable request.

\section*{AUTHOR DECLARATIONS}%
\subsection*{Conflict of Interest}
The authors have no conflicts to disclose.

\subsection*{Author Contributions}
{\bf{Y. Hirasaki}}: Conceptualization (equal); Formal analysis (lead); Investigation (lead); Methodology (lead); Software(lead); Validation (equal); Writing – original draft (lead).
{\bf{S. Daimon}}: Conceptualization (lead); Funding acquisition (equal); Investigation (supporting); Methodology (supporting); Project administration (lead); Software(equal); Supervision (supporting); Validation (equal); Writing – review \& editing (supporting).
{\bf{T. Itoko}}: Methodology (supporting); Validation (supporting); Writing – review \& editing (supporting).
{\bf{N. Kanazawa}}: Project administration (supporting); Software(supporting); Supervision (supporting); Writing – review \& editing (supporting).
{\bf{E. Saitoh}}: Funding acquisition (lead); Project administration (equal); Supervision (lead); Validation (equal); Writing – review \& editing (lead).

\bibliography{references}

\end{document}